\documentclass[twocolumn,floats,aps]{revtex4}
\usepackage{times,amsmath,amsfonts,psfrag,latexsym,graphics}
\begin{document}
\title{First-Order Reversal Curves of the Magnetostructural Phase Transition in FeTe}
\author{M. K. Frampton$^1$, J. Crocker$^1$, D. A. Gilbert$^{1,2}$, N. Curro$^1$, Kai Liu$^1$, J. A. Schneeloch$^3$, G. D. Gu$^3$, and R. J. Zieve$^{1}$}
\affiliation{${}^1$ Physics Department, University of California, Davis, CA  95616, USA\\
${}^2$ National Institute of Standards and Technology, Gaithersburg, MD 20899, USA\\
${}^3$ Brookhaven National Laboratory, Upton, NY 11973, USA}
\begin{abstract}
We apply the first-order reversal curve (FORC) method, borrowed from studies of ferromagnetic materials, to the magneto-structural phase transition of Fe$_{1+y}$Te. FORC measurements reveal two features in the hysteretic phase transition, even in samples where traditional temperature measurements display only a single transition.  For Fe$_{1.13}$Te, the influence of magnetic field suggests that the main feature is primarily structural while a smaller, slightly higher-temperature transition is magnetic in origin.  By contrast Fe$_{1.03}$Te has a single transition which shows a uniform response to magnetic field, indicating a stronger coupling of the magnetic and structural phase transitions. We also introduce uniaxial stress, which spreads the distribution width without changing the underlying energy barrier of the transformation. The work shows how FORC can help disentangle the roles of the magnetic and structural phase transitions in FeTe.
\end{abstract}
\maketitle

\begin{center}
\bf{INTRODUCTION}
\end{center}

For decades it was well-established wisdom that superconductors could not be magnetic, and could not even contain a substantial concentration of magnetic impurities. Yet beginning in the 1980's unconventional superconductors such as perovskites \cite{Kloss16}, 115s \cite{Weng16}, and other heavy fermions \cite{Flouquet91} illustrated that in some cases magnetism could coexist with and even enable superconductivity. The trend culminated with the iron-based chalcogenides and pnictides, which claim among the highest superconducting transition temperatures of above 50 K despite containing the quintessential magnetic atom \cite{Si16}. Understanding the complex interplay among the many types of interactions in these compounds may ultimately lead to higher-temperature superconductors or devices based on their other collective behaviors.

Many chalcogenides and pnictides exhibit structural and magnetic phase transitions which are nearly or exactly simultaneous and occur well above any superconducting transition temperature. At even higher temperatures compounds in both families also develop nematicity \cite{Wang16}. These material properties and their coupling explicitly define the electron spin ordering and phonon coupling, and set the stage for the superconducting transition. The magnetic and structural ordering are particularly intertwined in Fe$_{1+y}$Te: For small $y$, the high-temperature phase, which is tetragonal and paramagnetic, undergoes a first-order transition to a bicollinear antiferromagnetic phase with monoclinic structure \cite{Li09}. However, for sufficiently large $y$, the low-temperature phase changes to orthorhombic with incommensurate helical magnetic order \cite{Bao09}. This apparently second-order transition violates the weak Lifshitz criterion \cite{Michelson78}, which may indicate a precursor spin-liquid state above the transition \cite{Materne15}. For an intermediate range near $y=0.11$, a series of phases emerges on cooling. One set of measurements suggest an initial orthorhombic distortion coinciding with incommensurate antiferromagnetism, and then at lower temperatures a transition to a monoclinic, bicollinear antiferromagnet \cite{Mizuguchi12, Koz13}. More recent work supports an even more complicated sequence: first a monoclinic distortion, then the onset of incommensurate antiferromagnetism, and finally a zigzag distortion of the monoclinic lattice as the magnetic ordering becomes commensurate \cite{Zaliznyak12, Fobes14}. The upper two transitions are apparently second-order, while the lowest transition is strongly first-order. The lowest-temperature transition may also coincide with yet another type of order, an electronically driven ferro-orbital ordering that alters the magnetism and produces a structural distortion \cite{Fobes14, Bishop16,Ku09}.

It is expected that both the magnetic ordering and lattice structure play crucial roles in enabling superconductivity in these materials. Here we probe the intimate details of the magnetic and structural phase transitions and the magneto-structural coupling using a first-order reversal curve (FORC) technique \cite{Mayergoyz, Pike99, Davies}. By a controlled sequence of temperature cycles, this technique sets the system in the middle of the phase transition, then measures its evolution as it is driven out of the transition under increasing temperatures. The details of this evolution clearly depend on the distribution of intrinsic properties and interactions within the sample; the FORC technique evaluates the evolution from several mixed-phase starting conditions to separately extract these details - a feat which is impossible with standard resistance versus temperature measurements. The FORC technique is traditionally applied to magnetic materials \cite{Pike, Dumas07, Kirby, Rotaru, Gilbert13, Dobrota, Gilbert15}, and is able to quantitatively extract details including the magnetization reversal mechanism, the anisotropy distribution, and the magnetic dipolar and exchange interactions. In this work, we extract the phase transition activation energy -- an analog of the anisotropy -- and the strain-based interaction energy. These details cannot be easily extracted from the simple single parameter measurements typically performed. In addition, the FORC technique allows us to explicitly separate hysteretic (first-order) transitions from second-order transitions in the same temperature regime. Thus, the FORC measurements yield an unprecedented, microscopic view of the phase evolution.
 
\begin{center}
\bf{EXPERIMENT}
\end{center}

Single crystal samples of Fe$_{1.03}$Te and Fe$_{1.13}$Te were fabricated by unidirectional solidification, following procedures discussed previously \cite{Wen11}. In-plane resistance measurements were performed using a four-probe constant current configuration. Uniaxial stress was applied using a non-magnetic stainless steel press with manganin foil manometer \cite{Frampton17}.

\begin{figure}[!htb]
\begin{center}
\scalebox{0.28}{\includegraphics{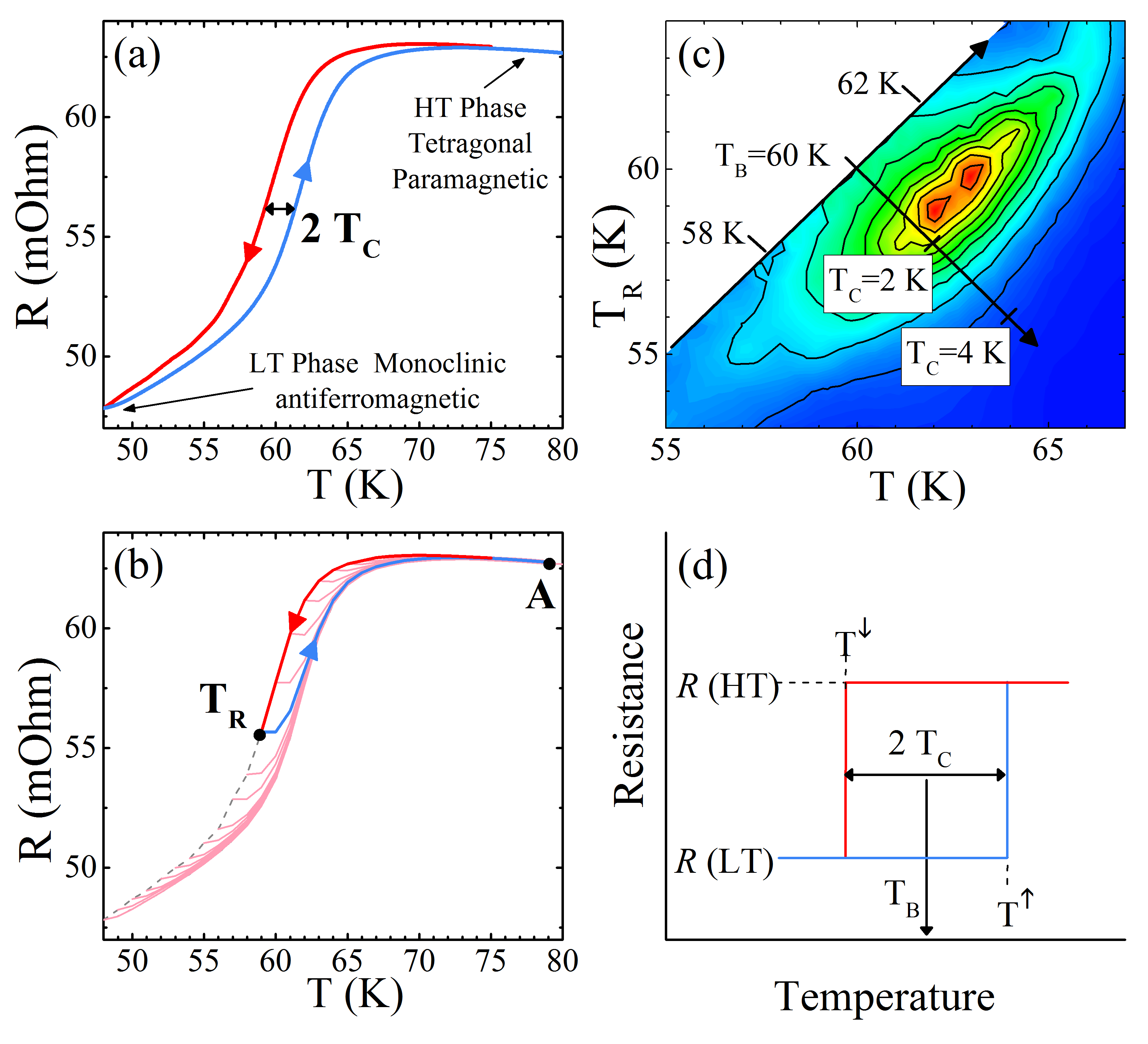}}
\caption{(a) Illustrative diagram of the warming and cooling lines demonstrating hysteresis and thermal coercivity, taken from Fe$_{1.03}$Te at H=0, P=120 Mpa. (b) the family of FORCs shown to fill the major loop, highlighting the FORC branch starting at $T_R$=59 K. (c) Calculated FORC distribution with ($T, T_R$) and ($T_C, T_B$) coordinate axes shown. (d) Schematic illustration of a hysteron associated with the thermal hysteresis.}
\label{f:FORCSchem}
\end{center}
\end{figure}

The transition which we monitor occurs near 62 K for Fe$_{1.03}$Te and 45 K for Fe$_{1.13}$Te. In each case the transition is thermally hysteretic; the resistances measured while cooling and warming do not overlap. The details of a hysteretic transition are directly related to the nanoscale properties of the system including the activation energy to initiate the transformation, the distribution of the activation energies across the sample, e.g. due to defects or strain fields, and the interactions between phases during the transformation. We will use ``intrinsic" to describe all influences that would affect the transition of a small isolated piece of the material, including such properties as activation energy or defect density.
Traditional measurements of a hysteretic transition start well away from the transition; we will take the high-temperature (HT) phase as the starting point. The resistance is measured as the temperature is decreased and the system transforms entirely into the low-temperature (LT) phase. The temperature sweep direction is then reversed, and the resistance is measured as the temperature increases and the system re-enters the HT phase. This forms the ``major hysteresis loop," which is shown for Fe$_{1.03}$Te at $H=0 T$, and $P=120$ MPa in Fig. \ref{f:FORCSchem}(a). The FORC technique obtains additional data corresponding to the interior of the major loop, by preparing the system in a mixed-phase state and measuring as it progresses towards a single-phase state. 

Our FORC measurement scheme is based on previously published procedures \cite{Gilbert13,Dobrota,Davies,Mayergoyz,Pike,Gilbert14,Ramirez}. As when measuring the major hysteresis loop, the sample is prepared in the HT phase, point $A$ in Fig. \ref{f:FORCSchem}(b). Then the temperature is lowered towards the LT phase (red curve). The temperature sweep is halted between the HT and LT single-phase states, at a temperature termed the ``reversal temperature" $T_R$. Between point $A$ and $T_R$ the resistance follows the major hysteresis loop, although these values are not used in the subsequent FORC analysis. The temperature is then increased in increments of $\Delta T$ from $T_R$ back towards $A$. At each temperature $T$, the resistance is measured after the sample achieves thermal equilibrium. A single FORC branch with $T_R$=59 K is highlighted in blue in Fig. \ref{f:FORCSchem}(b). Upon reaching $A$, the process is repeated for a new $T_R$, until these ``minor loops" fill the interior of the major loop; all of the measured FORCs, termed a family of FORCs, are shown in pink, with the major loop shown as dashed line in Fig. \ref{f:FORCSchem}(b). For $T_R$ near $A$ the sample is in the HT phase, and for $T_R$ at particularly low temperatures ($<50$ K in Fig. \ref{f:FORCSchem}) the sample is in the LT phase. Under our FORC procedure, for $T_R$ within the phase transition temperature range, the system will start in a mixed HT/LT state and end in the HT phase. The evolution of the system as $T$ increases from $T_R$ depends on the intrinsic and interaction details listed above. The temperature is changed quite slowly during these measurements, to ensure that there is no overshoot of $T_R$ or $T$. 

Once the family of FORCs is collected, the FORC distribution $\rho(T, T_R)$ is calculated by applying a mixed second order derivative, $\rho(T, T_R)=-\frac{\partial}{\partial T_R}(\frac{\partial R(T, T_R)}{\partial T})$, shown in Fig. \ref{f:FORCSchem}(c). The derivative $\frac{\partial R(T, T_R)}{\partial T}$ identifies the slope of R($T, T_R$), at each value of $T$; the subsequent derivative $\partial T_R$ identifies how the slope changes at a particular value of $T$ along branches starting at different $T_R$. Noting that the changes in the slope correspond to the physical LT-to-HT phase transitions, this sequence of measurements identifies changes in the transformation temperatures as a function of the phase state at each $T_R$.

The traditional approach to interpreting FORC distributions is to apply the Preisach model of hysteresis \cite{Stancu03}, which describes a hysteretic system as a weighted sum of fundamental units of hysteresis called hysterons. In this model each hysteron has one contribution to the resistance associated with the HT phase, and a different contribution associated with the LT phase, as illustrated in Fig. \ref{f:FORCSchem}(d). The hysteron has a sharp HT-to-LT transition temperature $T^\downarrow$ and a sharp LT-to-HT transition temperature $T^\uparrow$. In FORC measurements with $T_R > T^\downarrow$, the hysteron never leaves the HT phase; thus $\frac{\partial R(T, T_R)}{\partial T} = 0$ and is independent of $T_R$. Correspondingly the FORC distribution for this value of $T_R$ and $T$ is zero at all $T$, with the restriction that the measurements can only sample $T>T_R$. Similarly, for the FORC branches with $T_R < T^\downarrow$, the hysteretic element does reach the LT state. As the FORC measurement temperature increases, the sample remains in the LT state until $T= T^\uparrow$, at which point the hysteron switches back into the HT phase, and the resistance abruptly changes. The derivative $\frac{\partial R(T, T_R)}{\partial T}$ vanishes except at $T=T^\uparrow$, and since the measurements near $T^\uparrow$ are identical for all $T_R<T^\downarrow$ the mixed second derivative again vanishes. The only non-zero contribution to the FORC density $\rho(T,T_R)$ is at the crossover between these regimes, where $T_R=T^\downarrow$ and $T=T^\uparrow$. In this case varying either $T$ or $T_R$ switches between the two resistance levels. The entire sample is treated as a weighted sum of hysterons with unique ($T^\uparrow, T^\downarrow$) parameters, thus the FORC distribution maps out the weight parameter. Each hysteron encodes details of local behavior, which may be intrinsic or may stem from interactions.

Next, we define the center of a hysteron by $\frac{T^\uparrow + T^\downarrow }{2}$, and its coercivity by $\frac{T^\uparrow - T^\downarrow}{2}$. Bearing in mind that each hysteron's contribution to the FORC distribution appears at ($T=T^\uparrow, T_R = T^\downarrow$), a new coordinate system can be defined in terms of these values: ($T_C=\frac{T - T_R }{2}$, $T_B=\frac{T + T_R}{2}$) where $T_C$ and $T_B$ are the coercive and bias temperatures. The vocabulary stems from the original use of FORC with ferromagnets. The transformation of the FORC distribution is shown in Fig. \ref{f:FORCSchem}(c). Physically, $T_B$ is the average (center) of the phase transition temperatures, while $T_C$ identifies the energy barrier (or activation energy) of the phase transition. In particular, displacement of a feature along $T_C$ shows a change in the energy barrier for the transition, while displacement in $T_B$ shows a change in the center of the transition temperature. Any hysteretic transition has non-zero $T_C$, with larger values of $T_C$ corresponding to stronger hysteresis. At the other extreme, a second-order transition has no hysteresis, so it would appear along the $T_C=0$ axis. Without special consideration beyond the analysis described here, features do not appear on the $T_C=0$ axis \cite{Pike03}; thus the FORC technique filters out non-hysteretic transitions. The FORC diagrams in this text are plotted in the $(T, T_R)$ coordinates since these coordinates correspond to the temperatures set during measurement, and thus may be more simply understood. The $(T_C, T_B)$ coordinates are included in the plots, as they offer keen physical insight into the intimate details of the system.

For further insight into the physical origin of the FORC features, we consider, as an example case, a single very small isolated crystalline grain. If small enough, this one grain should exhibit sharp transitions and contribute to the FORC distribution only at $(T = T^\uparrow, T_R = T^\downarrow)$. Next consider a collection of isolated (non-interacting) small crystallites with slightly different transition temperatures, e.g. due to doping variation. If each crystallite has the same quality (defect density) the activation energy is expected to be the same, resulting in a common $T_C$. However, the different transition temperatures will displace the FORC features from each crystallite along the $T_B$ direction. Thus, the resultant feature will appear narrow in $T_C$ and elongated and continuous in the $T_B$ direction. As the difference between the transition temperatures becomes larger, the FORC features decouple and can become discrete features \cite{Gilbert14,Gilbert16}.Alternatively, consider a collection of crystallites with similar stoichiometric composition, but variations in their defect densities. The central transition temperatures $T_B$ for these crystallites should be the same, but the defects may act as nucleation sites for premature phase nucleation or pin the phase transition propagation front. The consequences of defects are symmetric along the warming and cooling branches, and hence will be manifested along the $T_C$ axis. Adding interactions to the example system significantly complicates the FORC distribution in non-trivial ways \cite{Gilbert13}, which are still under on-going investigation \cite{Gilbert14}. For this work, the relevant interaction is found to be a mean field-like destabilizing interaction. In this interaction the mixed-phase state is favored. As a consequence, the HT-to-LT transitions occurs prematurely along the cooling branch of the major loop from the HT phase to enter the mixed state. Similarly, the HT-to-LT transition is suppressed to lower temperatures when approaching the LT state on the cooling branch of the major loop to remain in the mixed state. These transitions act to expand the FORC distribution along the $T_R$ direction. Along each FORC branch, the same mean field interactions act to promote the mixed phase state, inducing similar shifts along the $T$-axis, resulting in a net shift along the $T_B$ direction, as discussed in Ref. \cite{Gilbert13}. As long as the intrinsic transition temperatures of the crystallites are close together, such an interaction is manifested as a broadening of the FORC distribution along the $T_B$ axis. Analogous discussion on mean-field interaction in FORC is provided in Ref. \cite{Gilbert13} and \cite{Dobrota}.

\begin{figure}[!htb]
\begin{center}
\scalebox{0.38}{\includegraphics{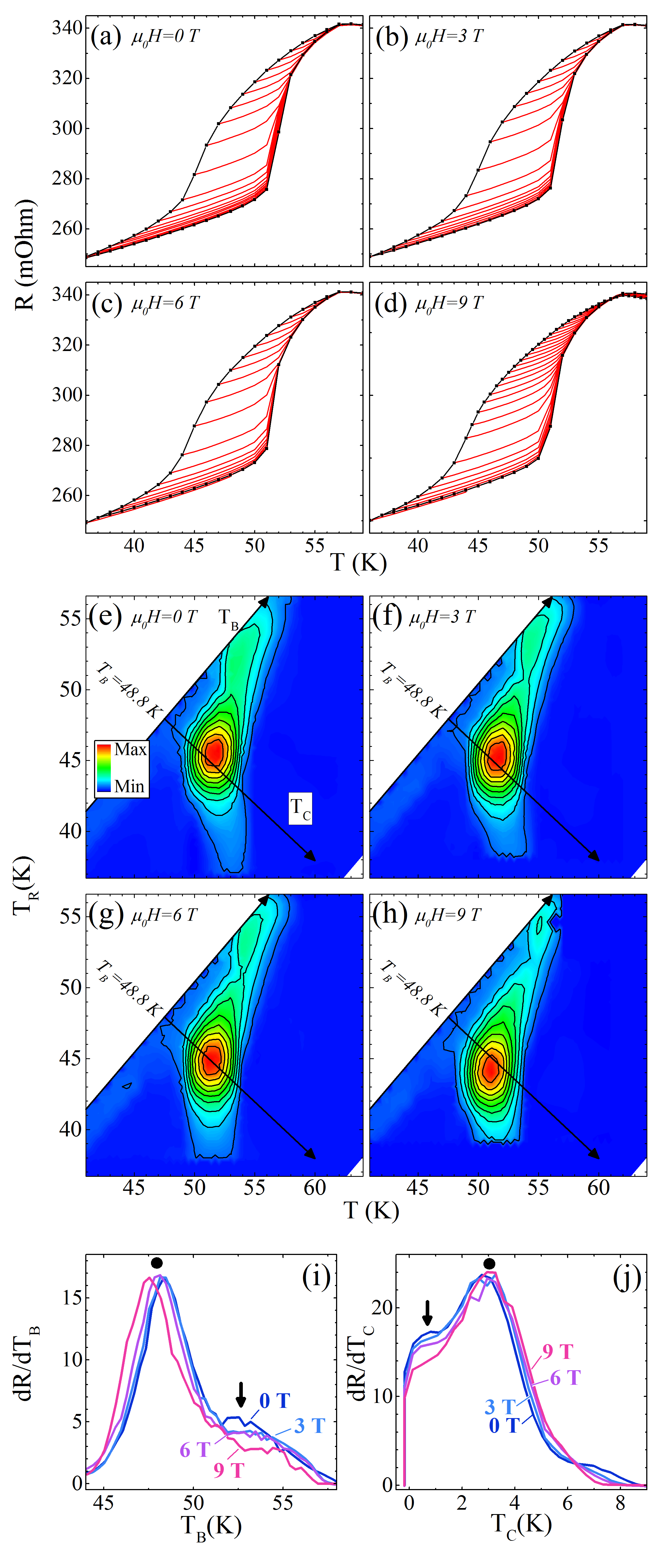}}
\caption{(a)-(d) Major hysteresis loop (black) and family of FORCs for Fe$_{1.13}$Te measured at P = 0 Pa and different magnetic fields applied along the $c$-axis. (e)-(h) FORC distributions extracted from (a)-(d). (i)-(j) Projections of (e)-(h) onto the $T_B$ and $T_C$ axes, normalized to their respective maxima, with the main peak and satellite feature indicated by a dot and arrow, respectively.  Error bars are determined by the resistance and temperature sensitivity, and are smaller than the line width.}
\label{f:field1.13}
\end{center}
\end{figure}

One other useful concept is the projection of the FORC distribution along one of its coordinates \cite{Dumas}. In some cases projections onto the $T$ or $T_R$ axes are informative; in others projections onto $T_C$ or $T_B$ are more useful. As seen in the discussion of hysterons, varying $T_R$ probes the HT to LT transition, so a projection onto the $T_R$ axis is essentially $\frac{\partial R(T, T_R)}{\partial T}$ of the major loop, measured upon cooling, but only identifying the hysteretic events. Similarly a projection onto the $T$ axis is $\frac{\partial R(T, T_R)}{\partial T_R}$ of the major loop, measured upon warming. However, projections onto $T_C$ and $T_B$ have no simple analogs in a measurement of only the major loop. Using the aforementioned description of $T_C$ and $T_B$ the FORC diagram in ($T_C, T_B$) coordinates plots the phase-resolved nanoscale transition temperature and the thermal coercivity - corresponding to the activation energy.

\begin{center}
\bf{RESULTS}
\end{center}

\begin{figure}[!htb]
\begin{center}
\scalebox{0.38}{\includegraphics{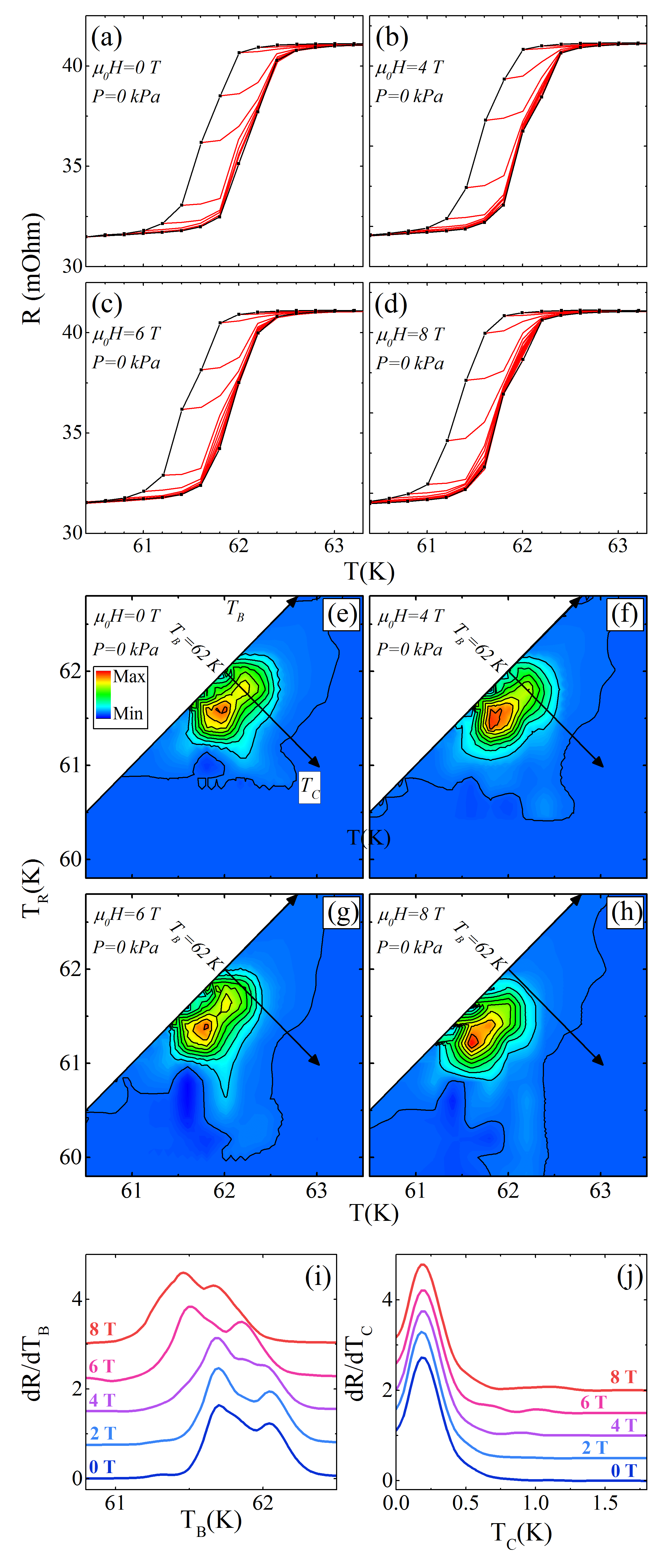}}
\caption{(a)-(d) Major hysteresis loop (black) and family of FORCs for Fe$_{1.03}$Te measured at P = 0 Pa and different magnetic fields applied along the $c$-axis. (e)-(h) FORC distributions extracted from (a)-(d). (i)-(j) Projections of (e)-(h) onto the $T_B$ and $T_C$ axes, with the main peak and satellite feature indicated by a dot and an arrow, respectively. The curves in each frame are normalized by the height of the main peak. The 2 T data are shown only in the projections, since the FORCs and resulting distribution are nearly identical to those for zero-field. Scaling and error bars are determined as in Fig. \ref{f:field1.13}. Plots in (i) and (f) are sequentially offset by 0.5 to improve visibility.}
\label{f:field1.03}
\end{center}\end{figure}

The measured family of FORCs and extracted FORC distributions for Fe$_{1.13}$Te at several magnetic fields are shown in Fig. \ref{f:field1.13}, panels (a)-(d) and (e)-(h), respectively. Different research groups report slightly different Fe content for the intermediate regime with multiple transitions \cite{Zaliznyak12, Koz13, Materne15}. Our Fe$_{1.13}$Te sample lies in that regime, with characteristic metallic behavior below the resistive transition \cite{Koz13}. Interestingly, the two-dimensional FORC contour plots show two distinct features: a dominant peak centered near ($T$=52 K, $T_R$=45 K) and a secondary peak at slightly higher $T$ and lower $T_R$ ($T$=54 K, $T_R$=54 K). The second peak indicates a two-step reversal and is entirely invisible in the major hysteresis loops, panels (a)-(d). In an applied magnetic field the main peak is shifted to lower $T$ and $T_R$ and the secondary satellite peak intensity is suppressed. The $T_B$ and $T_C$ projections, shown in Fig. \ref{f:field1.13}(i) and (j), confirm the displacement of the main peak in $T_B$ and suppression of the satellite peak with increasing magnetic field. The area of the satellite peak extrapolates to zero at 16.5 T, which is consistent with standard magnetic exchange coupling parameters, suggesting the origin of this peak may be local magnetic ordering. We note that the magnetic field has no influence on the width or position of the main peak in the $T_C$ projection, but only displaces the peak in $T_B$. This suggests a change to the transition temperature, but not the activation energy. As will be discussed below, similar trends are observed in the Fe$_{1.03}$Te sample. The uneven influence of field, suppressing the satellite peak but merely translating the main peak, suggests that both peaks possess a magnetic ordering component, but for the main peak the ordering may be directly coupled to a structural or orbital ordering which is not suppressed by the field. Additionally, the satellite peak occurs at a smaller value of $T_C$, generally corresponding to a transition with less hysteresis, and indicating a different (smaller) activation energy than the main peak. One possibility is that we are measuring the lower two of the three transitions identified by Fobes {\em et al.} \cite{Fobes14}. Our main peak is the orbital ordering transition which is accompanied by magnetic and electronic changes; this is consistent with the slight effect of magnetic field on this peak. The satellite peak, which appears to be magnetic, may be the antiferromagnetic transition previously identified as second-order. The FORC results suggest that it is in fact weakly first-order, to a degree difficult to ascertain with traditional transport measurements. 

\begin{figure}[!htb]
\begin{center}
\scalebox{0.38}{\includegraphics{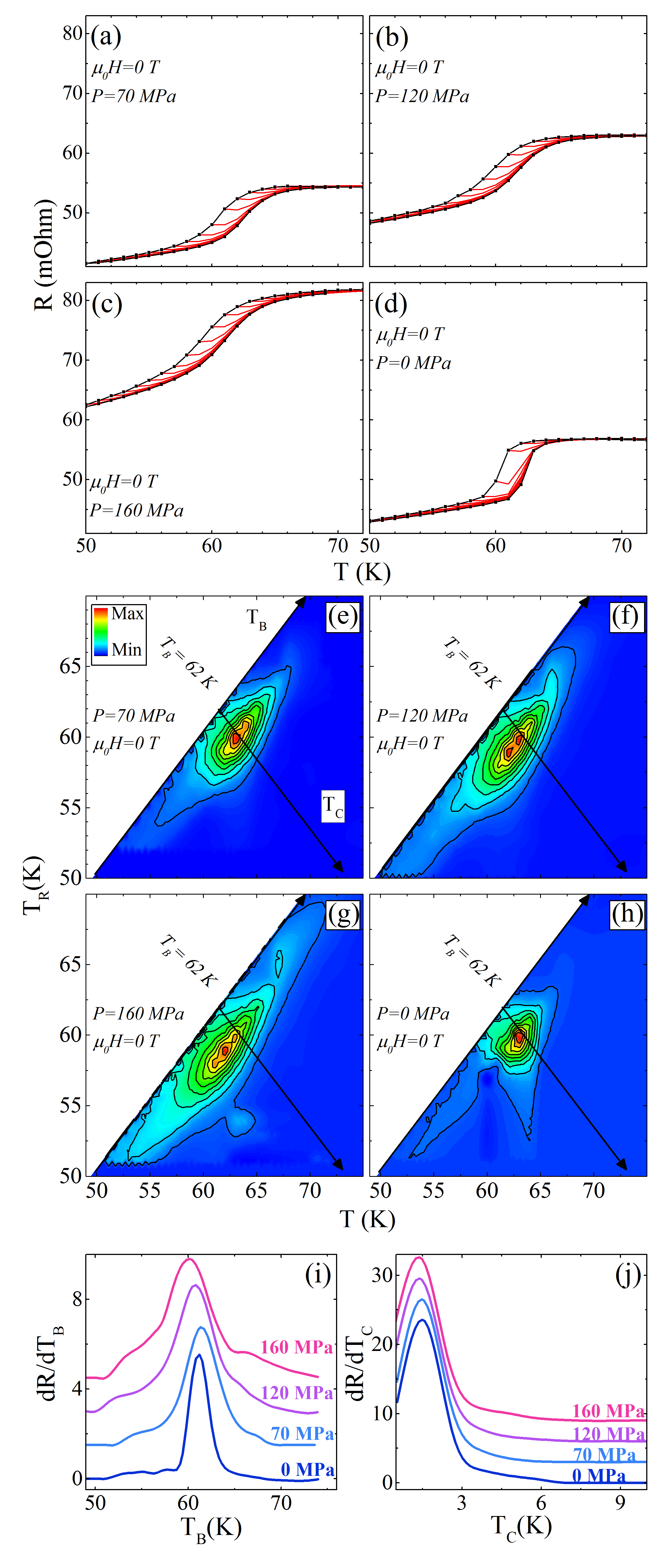}}
\caption{(a)-(d) Major hysteresis loop (black) and family of FORCs for Fe$_{1.03}$Te measured at $\mu_0H = 0$ T, at (a) $P = 70$ MPa, (b) $P = 120$ MPa, (c) $P = 160$ MPa, (d) $P = 0$ (order shows measurement sequence). (e)-(h) FORC distributions extracted from (a)-(d). (i)-(j) Projections of (e)-(h) onto the $T_B$ and $T_C$ axes, with the main peak and satellite feature indicated by a dot and an arrow, respectively. The curves in each frame are normalized by the height of the main peak. Error bars are determined by the resistance and temperature sensitivity, and are smaller than the line width. Scaling, offset and error bars are determined as in Fig. \ref{f:field1.03}.}
\label{f:pressure}
\end{center}
\end{figure}

Fig. \ref{f:field1.03} displays corresponding measurements on Fe$_{1.03}$Te, which has a monoclinic LT state. Here the zero-field FORC distribution is elongated in the $T_B$ direction and narrow in $T_C$. This is analogous to a magnetometry-FORC distribution for a sample with narrow intrinsic coercivity distribution and large mean-field-like demagnetizing interactions \cite{Gilbert13}. Demagnetizing interactions in magnetic materials destabilize the saturated magnetic state - or alternatively stabilize the demagnetized configuration. Translating this analogy to the FeTe FORCs, the elongation along the $T_B$ direction may indicate interactions which favor a mixed-phase state, or alternatively, destabilize the single-phase state. A narrow coercivity distribution is consistent with having a high-quality single crystal sample, while the origin of the destabilizing interactions is, as of yet, unclear. Another interpretation is that the two features may indicate different regions within the sample with slightly different transition temperatures, suggesting stoichiometry variation. As in Fe$_{1.13}$Te, while major hysteresis loops, panels (a)-(d), show what appears to be a single transition, the FORC distribution suggests a two-step phase evolution, as appears most clearly in the $T_B$ projection of Fig. \ref{f:field1.03}(e).

Upon the application of a magnetic field along the (001) axis, Fig. \ref{f:field1.03}(f-h), the FORC distribution does not exhibit appreciable deformation or changes in intensity, but again the feature is displaced in $-T_B$, particularly for $\mu_0H > 4$ T. This suggests that there is a magnetic component to the transition, but suppression of the magnetic ordering does not suppress other transformations - presumably orbital ordering.\cite{Fobes14} In the $T_B$ projection shown in Fig. 3(i), the two-peak feature does not move for $\mu_0H = 2$ T, then shifts steadily towards smaller $T_B$ for $\mu_0H = 4$ T - 8 T. Weighing the thermal energy and magnetic Zeeman energy against each other, an energy can be extracted from the shift, as shown in the discussion section. Similar to the main peak in the Fe$_{1.13}$Te sample, the $T_C$ projection, Fig. \ref{f:field1.03}(j), shows no dependence whatsoever on the magnetic field.

The above FORC measurements demonstrate magnetic field control of these
magneto-structural transitions, and suggest some coupling between
the magnetic ordering and  other transformations (structural and
orbital). Another approach to tuning the transitions is the application
of uniaxial stress; FORC measurements on Fe$_{1.03}$Te under uniaxial
stress are shown in Fig. \ref{f:pressure}. Stress is applied along the
(001) axis, without any external magnetic field. Unlike the family of
FORCs in Fig. \ref{f:field1.13} and Fig. \ref{f:field1.03} the pressure
measurements in Fig. \ref{f:pressure} generate a shift in the resistance
of both the HT and LT phase, shown in panels (a)-(d). However, the
derivative in the FORC calculation removes these offsets, only identifying
changes in the evolution processes. The extracted FORC distributions,
Fig. \ref{f:pressure}(e)-(h), show a distribution narrow in $T_C$ and
broad in $T_B$. Under increasing stress, panels (e)-(g), the width of
the FORC feature increases substantially in both directions of $T_B$,
but the distribution remains centered slightly below the $T_B = 62$
K line. Similar to the magnetic field case, the $T_C$ projection is
insensitive to pressure. Relieving the pressure, Fig. \ref{f:pressure}(h),
returns the FORC feature to a narrower distribution in $T_B$, suggesting
that the stress-induced changes are mostly reversible. A shallow tail
along $T_B$ which remains may be the result of residual sample damage
from the pressure cell.

\begin{center}
\bf{DISCUSSION}
\end{center}

The FORC distributions in Fig. \ref{f:field1.13} and Fig. \ref{f:field1.03} show that, above a critical field, the magnetic field uniformly translates the FORC distribution along $-T_B$, without changing its internal structure, as would be expected from Zeeman energy considerations. In an applied field, the larger magnetic susceptibility of the paramagnetic (HT) phase \cite{Koz13} stabilizes that phase relative to the antiferromagnetic (LT) phase. The Zeeman energy difference between the two phases, favoring the HT phase, shifts the transition to lower temperatures with increasing field. This holds for both directions of the transition, either to (HT-to-LT) or from (LT-to-HT) the antiferromagnetic phase; the entire hysteretic transition, along with its nesting behavior probed by FORC, moves to lower temperature. Since $T_C$ corresponds roughly to the half-width of the hysteresis loop, and is insensitive to the average transition temperature, the absence of a change in $T_C$ confirms exactly this behavior.

\begin{figure}[!htb]
\begin{center}
\scalebox{0.21}{\includegraphics{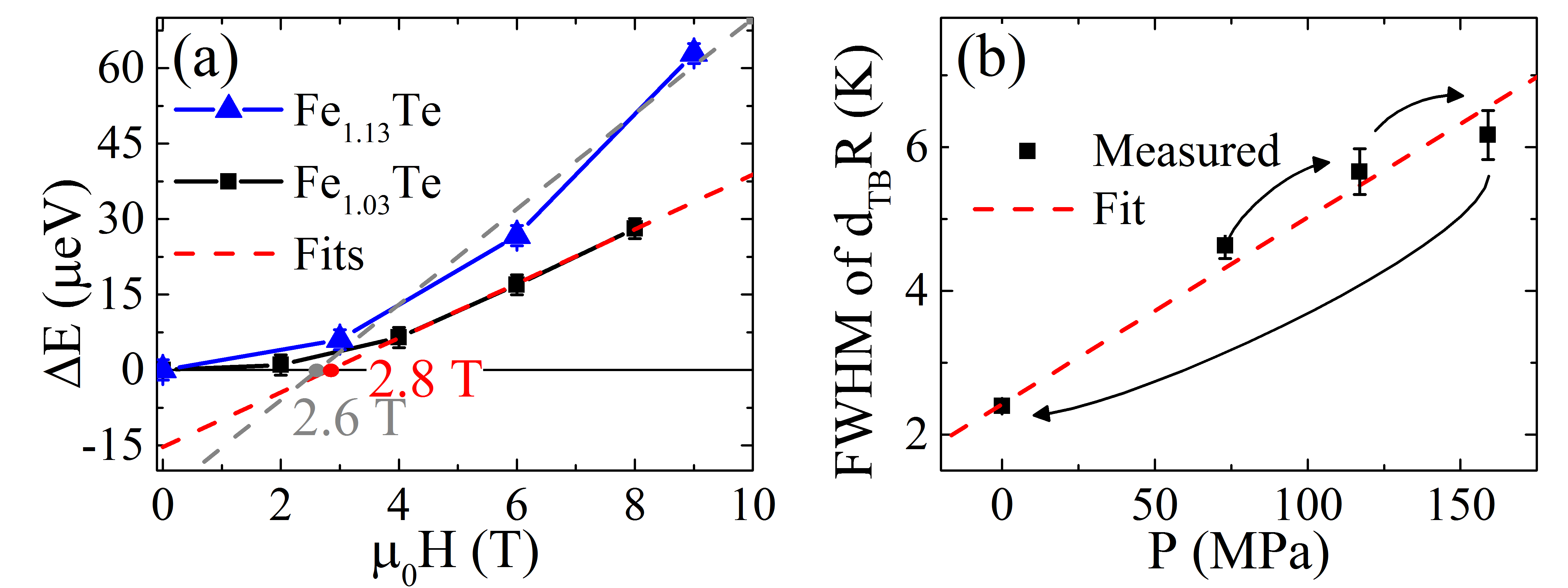}}
\caption{(a) Thermal energy calculated from the FORC feature displacement, as observed in the $T_B$ projections. Linear fit is calculated from the $\mu_0H > 2$ T data. (b) FWHM of FORC distribution under increasing pressure. Arrows indicate measurement sequence. Linear fit is determined using all of the data. Error bars are determined by the error of the peak fit location and width for (a) and (b), respectively.}
\label{f:energy}
\end{center}
\end{figure}

However, the traditional Zeeman energy is linear in applied field, implying the displacement of the FORC features would also be linear in field. We plot the $T_B$ displacement of the FORC features for the Fe$_{1.13}$Te and Fe$_{1.03}$Te samples in Fig. \ref{f:energy}(a). This plot shows the shift in $T_B$ of the fitted Gaussian center of $dR/dT_B$ between $H=0$ and an applied field $\mu_0H$, multiplied by Boltzmann's constant to convert to the corresponding energy. In both cases the displacement at low fields does not follow the expected linear trend. Indeed, the shift for $\mu_0H < 3$ T  is extremely small, as shown directly in Fig. \ref{f:field1.13}(e) and Fig. \ref{f:field1.03}(e), respectively. We conclude that orbital ordering energy considerations dominate the low-field phase transition, with magnetic contributions playing only a small role. Assuming a linear coupling between the energy and magnetic field, as is the case with the Zeeman interaction, we can quantify this statement; linearly extrapolating the FORC displacement from the $\mu_0H > 3$ T range, the magnetic field energy and other ordering energies (structural, orbital) become equal at 2.6 T and 2.8 T for Fe$_{1.13}$Te and Fe$_{1.03}$Te, respectively. A further extrapolation to zero magnetic field yields a coupling energy of 25 $\mu$eV for Fe$_{1.13}$Te and 15 $\mu$eV for Fe$_{1.03}$Te. It is important to note that these results were made possible by resolving the energy difference represented in the $T_B$ coordinate, which could be achieved only through FORC analysis.

The broadening under stress is plotted in Fig. \ref{f:energy}(b), which shows the full width at half-maximum of Gaussian fits to the $T_B$ projections of Fig. \ref{f:pressure}(i). The broadening both to higher $T_B$ and lower $T_B$ means that some regions in the sample transition at higher temperatures and others at lower temperatures, relative to their unstressed state. Meanwhile, the stability of the distribution in $T_C$, Fig. \ref{f:pressure}(j), implies that each region retains its original degree of hysteresis. The symmetrical broadening leads to the seemingly paradoxical conclusion that stress has opposite effect on the regions of the sample near the high or low ends of the transition, moving the former to yet higher temperature while depressing the latter. 

As noted above, the elongated feature suggests interactions which promote instability in the single-phase states \cite{Kou,Gilbert13,Dobrota,Valcu}. The pressure induced elongation of the FORC feature supports a conclusion that self-destabilizing interaction cause the stretching of the feature, rather than stoichiometric variation. One possible mechanism is that local strains created in either the tetragonal or monoclinic phases could be relieved in a mixed state with regions of each structure. Such a mechanism that favored the mixed state would spread out the transition as observed. To achieve the fully LT phase would require overcoming this additional energy and suppress the FORC feature to lower $T_R$. Subsequently warming from the LT state, the transition would begin at lower temperatures to return to the mixed-phase state, leading to the FORC feature at the same $T_C$, but displaced in $T_B$. Similarly, this preference for a mixed phase state causes the initial phase transitions to occur at higher temperatures, stretching the FORC distribution to a higher $T_B$. This type of interaction need not change the energy of the mixed-phase state, so the center of the FORC distribution could remain unchanged, as observed here. The broad uniformity of the FORC distribution along $T_B$ indicates that these interactions are mean-field-like, as opposed to local interactions, which would manifest as discrete maxima in the FORC distribution. For completeness, if the system favored a single-phase state, the transformation would occur as an avalanche event, resulting in a collapse of the FORC distribution feature to a single point \cite{Gilbert13}.

\begin{figure}[!htb]
\begin{center}
\scalebox{0.25}{\includegraphics{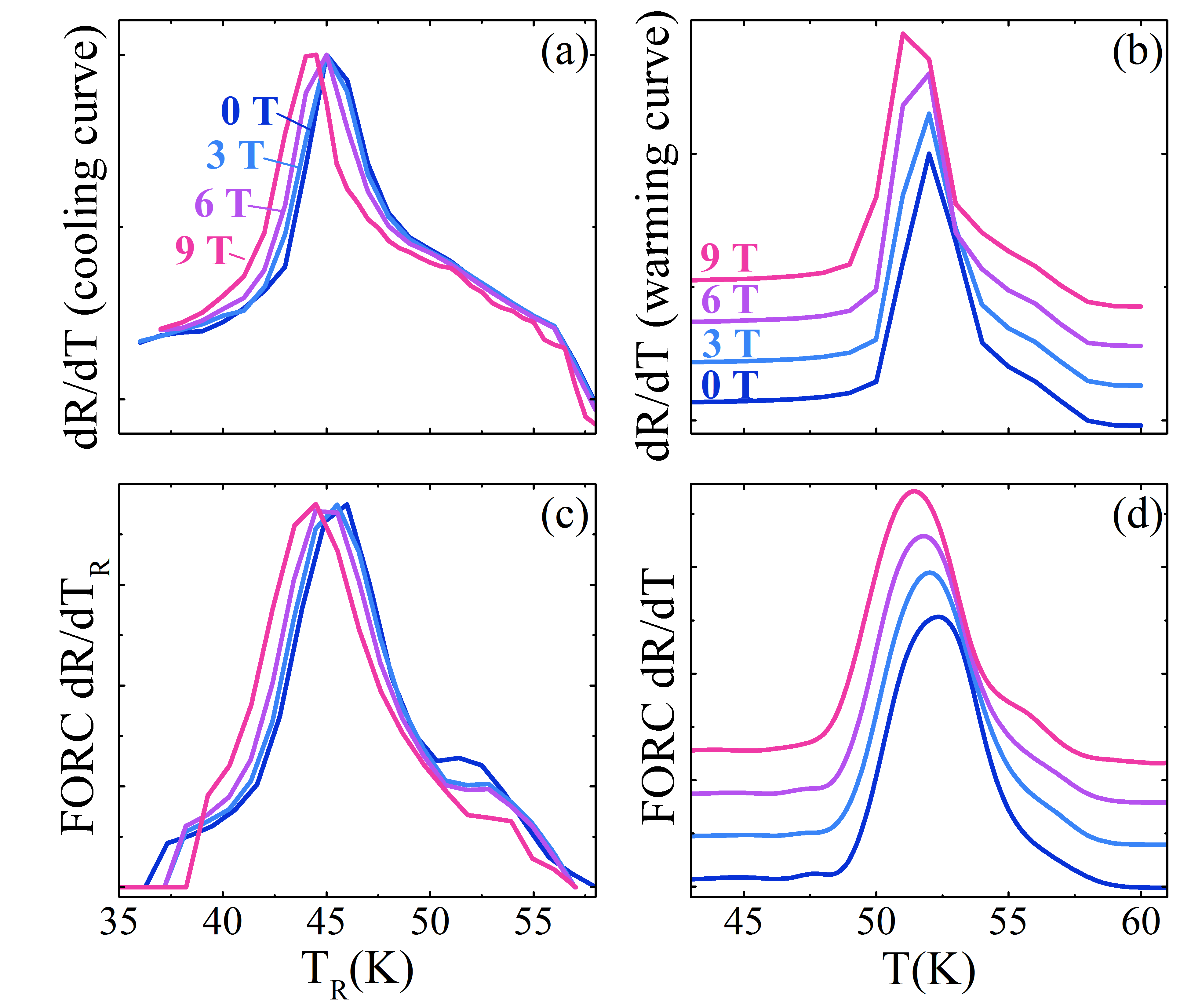}}
\caption{Comparison of (a, b) the derivative of the cooling and warming ``direct" measurements with (c, d) the $T_R$ and $T$ FORC projections from Fig. \ref{f:FORCSchem}(a-d) in the main text.}
\label{f:compare}
\end{center}
\end{figure}

Since the FORC technique is new to this research field,
Fig. \ref{f:compare} shows a comparison of the FORC projections
with the derivative of the warming and cooling one-dimensional
measurements. Specifically, the FORC measurements require a number of
minor loop measurements, followed by a complex mathematical operation to
develop a two-dimensional contour plot of the transitions. The advantage
of the FORC technique is its ability to quantitatively resolve both the
$T^\uparrow$ and $T^\downarrow$ events for the transitions, but throughout
this text the projections were provided as a powerful tool to aid in
visual clarification. Projecting the FORC distribution in either $T$ or
$T_R$ parameters returns the FORC measurement to a one-dimensional plot
which represents the sum total of all transitions along the warming
and cooling branches. Thus, the FORC projection should be similar
to the first derivative of the traditional one-dimensional warming
and cooling measurements, with a main difference being that the FORC
measurement shows only hysteretic transitions. In Fig. \ref{f:compare}
the derivative of the cooling and warming major loop measurements,
panels (a) and (b), and the FORC projections, panels (c) and (d), for
Fig. \ref{f:field1.13} are compared. The similarities between the curves
are immediately apparent, with all of the plots showing a shift of the
main peak to lower temperatures with increasing magnetic field. However,
there are also key differences, notably the derivative of the cooling
curve, panel (a), identifies a broad transition before the main peak, but
does not identify it as a separate, distinguishable transition, and does
not show a change in intensity. This highlights the ability of the FORC
technique to specifically resolve hysteretic transitions. Interestingly,
the feature corresponding to the suppressed high-temperature transition
shows only a change in intensity for the FORC measurement, implying the
magnetic field makes this transition more reversible. Additionally,
the projections show a shift in both $T^\uparrow$ and $T^\downarrow$
(implicit in $T$ and $T_R$) but it is only by transforming into ($T_C$,
$T_B$) coordinates - which have no analogue in the direct measurement -
that we can directly relate to nanoscale physics effects.

\begin{center} \bf{CONCLUSIONS} \end{center}

In summary, the first order reversal curve (FORC) technique is applied to the magnetostructural phase transition of FeTe. The phase transition at $\approx$60 K is of particular interest as it sets the stage for the superconducting transition, and engenders the onset of magnetostructural coupling. The roles of stoichiometric doping and pressure are investigated, as these approaches are often used to induce superconductivity in this class of material. FORC measurements on FeTe show a two-step hysteretic transition, while traditional measurements show only a single first-order transition. In Fe$_{1.13}$Te one transition is suppressed by magnetic fields and seems to be the antiferromagnetic ordering previously thought to be second-order. This possibility of a first-order transition is notable because the order of the transition helps determine what phases are present elsewhere in the phase diagram. In Fe$_{1.03}$Te, low magnetic fields do not affect the phase transition, while larger fields cause a linear shift in the transition temperature, consistent with a simple Zeeman energy. Using this model, the magnetostructural coupling energy is quantitatively determined. Lastly, the FORC technique reveals that pressure increases the spread of the transition temperature without changing the activation energy. By analogy to magnetic FORC measurements, the measured FORC distribution implies a self interaction which stabilizes the mixed-phase state. The FORC technique thus provides unprecedented insight into the magnetostructural coupling and the consequences of stoichiometric doping and pressure, crucial to expanding our understanding of these materials. 

\begin{center}
\bf{ACKNOWLEDGEMENTS}
\end{center}

The authors acknowledge support from NSF through DMR-1609855 (R.J.Z),
DMR-1506961 (J.C. and N.C.), DMR-1008791 (D.A.G.), and ECCS-1611424
(K.L.). D.A.G. also acknowledges support from the NRC RAP.  The work at
Brookhaven National Laboratory was supported by the Office of Basic Energy
Sciences, DOE, under Contract No. DE-SC00112704. J.A.S. was supported by
the Center for Emergent Superconductivity, and Energy Frontier Research
Center funded by the Office of Basic Energy Sciences, DOE.  \\

\end{document}